\documentclass[a4paper,fleqn,usenatbib,useAMS,usedcolumn]{mnras}

\usepackage[T1]{fontenc}
\usepackage{ae,aecompl}

\usepackage{etoolbox}
\makeatletter
\patchcmd\@combinedblfloats{\box\@outputbox}{\unvbox\@outputbox}{}{%
 \errmessage{\noexpand\@combinedblfloats could not be patched}%
}%
\makeatother

\usepackage{graphicx}	
\usepackage[fleqn]{amsmath}	
\usepackage{amssymb}	

\usepackage{hyperref} 
\hypersetup{
    colorlinks   = true,
     citecolor    = blue
}

\newcommand{\sub}[1]{\ensuremath{_\mathrm{#1}}}
\newcommand{\super}[1]{\ensuremath{^\mathrm{#1}}}

\newcommand{\Msun}{\ensuremath{\mathrm{M}_\odot}}
\newcommand{\Mc}{\ensuremath{\mathcal{M}}}
\newcommand{\Mtot}{\ensuremath{M_\mathrm{total}}}

\newcommand{\dd}{\ensuremath{\mathrm{d}}}

\newcommand{\RNum}[1]{\uppercase\expandafter{\romannumeral #1\relax}}

\title[Measuring IMRACs with GWs]{Inference on gravitational waves from coalescences of stellar-mass compact objects and intermediate-mass black holes}

\author[C.-J. Haster et al.]{\parbox{\textwidth}{
Carl-Johan Haster$^{1},$\thanks{E-mail: cjhaster@star.sr.bham.ac.uk (CJH)}
Zhilu Wang$^{2, 1}$,
Christopher P.\ L.\ Berry$^{1}$,
Simon Stevenson$^{1}$,
John Veitch$^{1}$
and Ilya Mandel$^{1}$}
\vspace{0.2cm}\\
\parbox{\textwidth}{$^{1}$School of Physics and Astronomy, University of Birmingham, Edgbaston, Birmingham B15 2TT, United Kingdom}\vspace{0.2cm}\\
\parbox{\textwidth}{$^{2}$Department of Modern Physics, University of Science and Technology of China, 96 JinZhai Road, Baohe District, Hefei, Anhui, 230026, People's Republic of China}
}

\date{\today}

\pubyear{2015}

\begin{document}
\label{firstpage}
\pagerange{\pageref{firstpage}--\pageref{lastpage}}
\maketitle

\begin{abstract}
Gravitational waves from coalescences of neutron stars or stellar-mass black holes into intermediate-mass black holes (IMBHs) of $\gtrsim 100$ solar masses represent one of the exciting possible sources for advanced gravitational-wave detectors. These sources can provide definitive evidence for the existence of IMBHs, probe globular-cluster dynamics, and potentially serve as tests of general relativity. We analyse the accuracy with which we can measure the masses and spins of the IMBH and its companion in intermediate-mass ratio coalescences. We find that we can identify an IMBH with a mass above $100~\Msun$ with $95\%$ confidence provided the massive body exceeds $130~\Msun$. For source masses above $\sim 200~\Msun$, the best measured parameter is the frequency of the quasi-normal ringdown. Consequently, the total mass is measured better than the chirp mass for massive binaries, but the total mass is still partly degenerate with spin, which cannot be accurately measured. Low-frequency detector sensitivity is particularly important for massive sources, since sensitivity to the inspiral phase is critical for measuring the mass of the stellar-mass companion. We show that we can accurately infer source parameters for cosmologically redshifted signals by applying appropriate corrections. We investigate the impact of uncertainty in the model gravitational waveforms and conclude that our main results are likely robust to systematics.
\end{abstract}




\hypersetup{linkcolor=black}

\section{Introduction} 
\label{sec:intro}

The Advanced LIGO \citep[aLIGO;][]{TheLIGOScientific:2014jea} gravitational-wave (GW) detectors began their first observing run on 18 September 2015; the Advanced Virgo \citep[AdV;][]{TheVirgo:2014hva} GW detector is expected to commence scientific observation in 2016 \citep{Aasi:2013wya}.
One of the key sources for the advanced-era detectors are compact binary coalescences \citep[CBCs;][]{Abadie:2010cf}, the inspiral and merger of binary systems including both neutron-star (NS) and black-hole (BH) companions across a large mass spectrum.

Intermediate mass black holes (IMBHs) fill the gap in the continuum between stellar-mass BHs and supermassive BHs, potentially representing an early stage in the evolution of supermassive BHs \citep{Miller:2003sc,Graham:2012pw}. At present, the best evidence for their existence comes from observations of ultraluminous X-ray sources \citep{Feng:2011pc,Pasham:2015tca}.

IMBHs of a few hundred solar masses in a coalescing binary are a potential source of GWs for the advanced generation of ground-based detectors. If the IMBH's companion is another IMBH, the system is referred to as an IMBH binary (IMBHB). If its companion is stellar mass, then the system undergoes an intermediate mass-ratio coalescence (IMRAC). These are often also referred to as intermediate-mass-ratio inspirals (IMRIs);\footnote{The inspiral of an IMBH into a supermassive BH is also referred to as an IMRI. GWs from such IMRIs are potential sources for a space-borne detector \citep{AmaroSeoane:2007aw}, as are the most massive (redshifted total masses of $\gtrsim 10^3~\Msun$) IMBHBs \citep{Fregeau:2006yz,Miller:2008fi}.} however, we prefer IMRAC to highlight the importance of the entire coalescence, including the merger and ringdown phases in addition to the inspiral, to the detection and analysis of these high-mass systems \citep{Smith:2013mfa}.

IMRACs are most likely to be found in the dense cores of globular clusters \citep{Leigh:2014oda,MacLeod:2015bpa}. They can form through a range of mechanisms, including hardening of an existing binary, through either three-body interactions or Kozai oscillations as part of a hierarchical system, as well as through direct or tidal capture \citep{Mandel:2007hi}. As a consequence of mass segregation in globular clusters, the stellar-mass companion to the IMBH will change throughout the evolutionary history of the cluster \citep{MacLeod:2015bpa}. Soon after the formation of the cluster, the companion will most likely be a stellar-mass BH, but for older clusters it could be a NS if the stellar-mass BH population has been depleted by mergers and dynamical ejections \citep{Gill:2008pt, Umbreit:2012vx, Morscher:2014doa}. Consequently, there is a large variation in the possible mass ratios of detectable binaries.

The resulting IMRACs are estimated to have become largely circularized before entering the sensitivity band of the advanced GW detectors, and any residual eccentricity is expected to have a negligible effect on their detectability \citep{Mandel:2007hi}. Estimates of the IMRAC detection rate in the advanced-detector era range up to tens of events per year, though rates of zero are possible given the absence of confirmed IMBHs in the few-hundred-solar mass range where advanced detectors would be sensitive to emitted GWs \citep{Brown:2006pj,Mandel:2007hi,Abadie:2010cf}.\footnote{For comparison, coalescence rates of stellar-mass BH binaries originating in globular clusters could be $\sim100$ per year \citep{Rodriguez:2015oxa}.}

The dividing line between IMBHBs and IMRACs, just like the division between IMBHs and stellar-mass BHs, is arbitrary; however, the evolution of the binary and the emitted GWs vary significantly with the mass ratio $q=m_2/m_1$, where $m_1>m_2$ are the masses of the binary companions. Systems with more equal masses (IMBHBs with $q \sim 1$) inspiral more quickly than those with unequal masses (IMRACs with $q \ll 1$), and because of the difference in the scales associated with unequal masses, IMRACs are more challenging for numerical relativity to simulate \citep{Lousto:2010tb,Husa:2015iqa}.

We perform a systematic parameter-estimation (PE) study for IMRAC signals using full inspiral--merger--ringdown waveforms. Details of our set up, which mirrors the analysis of \citet{Veitch:2015ela} for IMBHBs, are described in \autoref{sec:PE}. Results are given in \autoref{sec:key_results} and discussed in \autoref{sec:discussion}, where we also examine the sensitivity of our analysis to systematics. Our main conclusions are summarized in \autoref{sec:summary}; in particular, we find that the advanced GW detectors could unambiguously confirm the existence of IMBHs should a suitable IMRAC ($m_1 \gtrsim 130~\Msun$) be detected.

\section{Study design}
\label{sec:PE}

To determine how accurately properties of IMRACs could be measured in the advanced-detector era, we analysed a mock set of GW signals observed with aLIGO and AdV. The simulated GW signals (injections) and the detector properties are described in \autoref{sec:injections}, and the data analysis is detailed in \autoref{sec:PE_summary}.

\subsection{Sources and sensitivity}
\label{sec:injections}

Following \citet{Veitch:2015ela}, we simulated a set of IMRAC signals which systematically cover the mass range of interest. We set the total mass of the binary $\Mtot=m_1+m_2$ to be between $50~\Msun$ and $350~\Msun$. For each total mass, three mass ratios $q$ of $1/15$, $1/30$ and $1/50$ were used.

Each injection was assigned a sky position isotropically drawn from the celestial sphere, as well as an inclination $\iota$ drawn from a distribution uniform in $\cos\iota$. The distance to each source was then selected to yield a signal-to-noise ratio (SNR) of $\rho = 15$, distributed across the detector network, in order to give an indication of typical PE accuracy; the dependence of PE on $\rho$ is investigated in \autoref{sec:SNR}.\footnote{The median SNR of detected signals, assuming that sources are uniformly distributed in (Euclidean) volume is $\rho\sub{med} = 2^{1/3} \rho\sub{det}$, where $\rho\sub{det}$ is the detection threshold \citep{Schutz:2011tw}. Taking a detection threshold of $\rho\sub{det} = 12$ \citep{Aasi:2013wya}, $\rho\sub{med} \simeq 15$.}

The injected signals were generated using a spin-aligned effective-one-body--numerical relativity (SEOBNR) waveform approximant, specifically SEOBNRv2 \citep{Taracchini:2012ig,Taracchini:2013rva}. These waveforms are constructed via the effective-one-body formalism \citep{Buonanno:1998gg,Buonanno:2000ef} for the inspiral dynamics, with the merger and ringdown portions calibrated to a suite of numerical relativity simulations \citep[e.g.,][]{Mroue:2013xna}. The companion spins are assumed to be aligned; including effects of generic spin alignments (such that there is precession) is an area of active development \citep{Pan:2013rra}. We only inject signals from non-spinning systems for this first study; we hope to include full spin effects in the future.
Additionally, we assume the binaries to be fully circularized before they enter the detectors' sensitive frequency band; as waveform approximants allowing for eccentricity effects become available for PE studies, we hope to include them as well.

The output of GW detectors is the sum of the GW strain and random detector noise. The particular noise realisation present determines which GW template best matches the data. The best matching template may have parameters offset from the true value; over many different realisations of the noise, this shift in the parameter estimates should average to zero.\footnote{The presence of non-stationary noise features (glitches) could impact PE leading to systematic errors. Realistic non-stationary, non-Gaussian noise has been shown not to affect PE performance for binary neutron stars \citep{Berry:2014jja}; however, these noise features could be more significant in analysing short-duration, low-frequency IMRAC signals.} However, at a given $\rho$, the measurement uncertainty should not be significantly influenced by the details of noise realisation. Since we are primarily concerned with PE accuracy, we use zero-noise injections; this simplifies comparison between different simulations as we only need to consider differences in the input parameters and not the noise. 

We assume that aLIGO and AdV are operating at their respective design sensitivities \citep{PSD:aLIGO, AdvVirgo}, which are expected to be realised at the end of the decade \citep{Aasi:2013wya}. To fully utilise the detectors' sensitivity to IMRAC sources, a lower frequency cut-off of $f\sub{low} = 10~\mathrm{Hz}$ was chosen for all three detectors; the importance of this low-frequency sensitivity is examined in \autoref{sec:flow}. 

\subsection{Parameter estimation}
\label{sec:PE_summary}

The data, with injected signals, were analysed using the Bayesian PE pipeline \texttt{LALInference} \citep{Veitch:2014wba}.\footnote{A component of the LIGO Scientific Collaboration Algorithm Library (LAL) suite \href{http://www.lsc-group.phys.uwm.edu/lal}{http://www.lsc-group.phys.uwm.edu/lal}.} For each event, \texttt{LALInference} computes a set of samples drawn from the joint posterior probability distribution spanning the signal parameters. To calculate the posterior, we need a model for the likelihood and prior probability distributions for the parameters.

The likelihood is calculated by matching a template signal to the data \citep{Cutler:1994ys}. The analysis was performed using the \mbox{SEOBNRv2\_ROM\_DoubleSpin} waveform approximant \citep{Purrer:2014fza,Purrer:2016inprep}, a reduced-order model (ROM) surrogate of SEOBNRv2 implemented in the frequency domain. The development of this ROM has enabled PE studies previously deemed computationally infeasible, expanding the accessible parameter space for studies of CBC sources \citep[cf.][]{Veitch:2015ela}. By using the same approximant for injection and recovery, we remove any systematic error caused by waveform uncertainty (cf.\ \autoref{sec:systematics}). SEOBNRv2 and its ROM surrogate only include the leading order quadrupolar mode of the GW radiation, but as it has been shown that higher-order modes can significantly improve the PE for IMBHB systems \citep{Graff:2015bba}, we hope to be able to include them in future IMRAC studies. 

For this analysis, we adopted a flat prior distribution on the companion masses $m_1,\,m_2 \in [0.6, 500]~\Msun$ with the constraints $\Mtot >12~\Msun$ and $q>0.01$. While all injections were non-spinning, we do allow for the exploration of dimensionless spin magnitudes $a_1,a_2 \in [-1, 0.99]$, aligned with the orbital angular momentum, with uniform priors. We also assume an isotropic prior on the source position and orientation in the sky as well as a uniform-in-volume prior on the luminosity distance out to $4~\mathrm{Gpc}$.

GWs are redshifted in an expanding Universe. This corresponds to redshifting all masses from a source at redshift $z$ by a factor of $(1+z)$, and scaling the GW amplitude with the inverse of the luminosity distance. In this study, we report the injected and recovered masses as redshifted to the detector rest frame, rather than the physical source frame masses, except where otherwise noted. The implications of cosmological effects are discussed in further detail in \autoref{sec:Cosmo}.

\section{Key results}
\label{sec:key_results}

We characterize the posterior probability distributions produced by \texttt{LALInference} in terms of the innermost $90\%$ credible intervals $\mathrm{CI}_{0.9}$, spanning the $5$th to the $95$th percentiles, for one-dimensional marginalized parameter distributions \citep{Aasi:2013jjl}.

\begin{figure}
\includegraphics{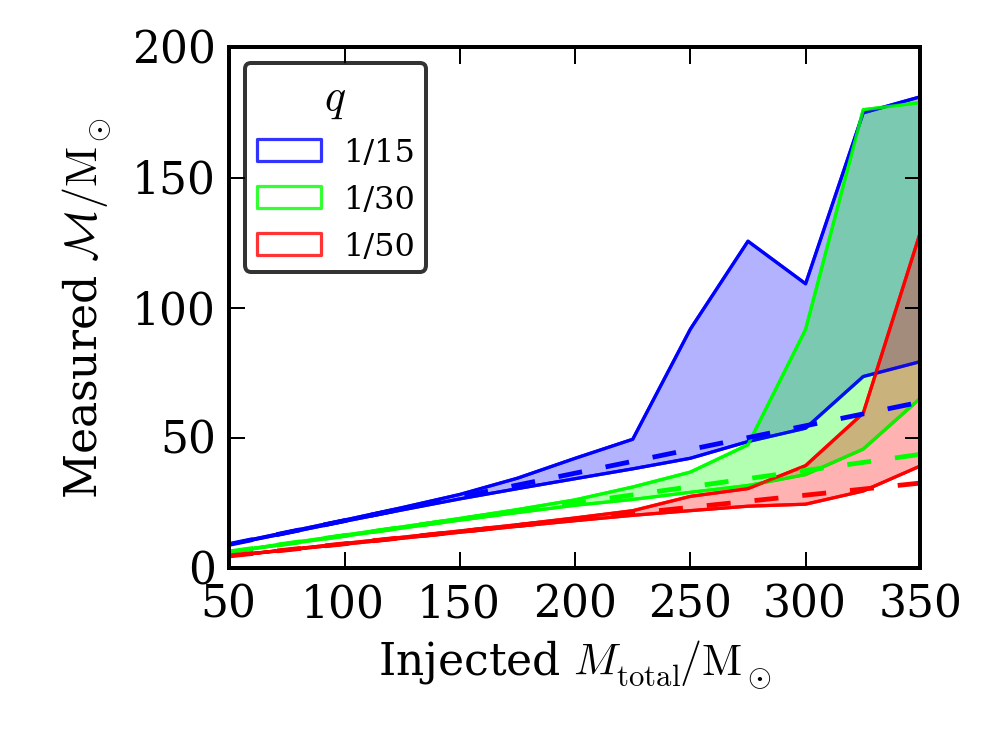}
\caption{The $90\%$ credible interval for the chirp mass $\Mc$ as a function of $\Mtot$ and $q$. The true values are shown as dashed lines for each $q$. The $\Mc$ measurement accuracy gradually worsens as $\Mtot$ increases up to $\Mtot \sim 200~\Msun$ and then deteriorates markedly in the region where little of inspiral phase falls into the sensitive frequency band of the detectors (cf.\ \autoref{fig:McMtot_frac}).}
\label{fig:Mc_bounds} 
\end{figure}

\begin{figure}
\includegraphics{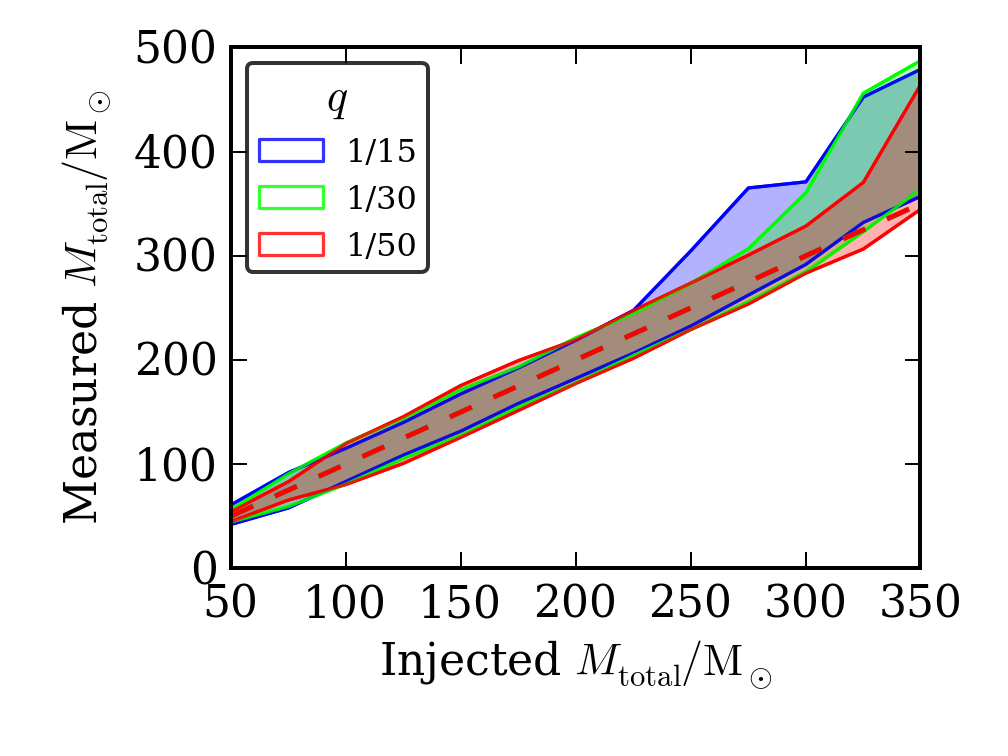} 
\caption{The $90\%$ credible interval for $\Mtot$. The true values are shown as a dashed line. For higher $\Mtot$ the $\mathrm{CI}_{0.9}$ widens, and is biased above the true value of $\Mtot$ as a result of the strong prior support for systems at higher luminosity distances (cf.\ \autoref{fig:components}).} 
\label{fig:Mtot_bounds} 
\end{figure}

\begin{figure}
\includegraphics{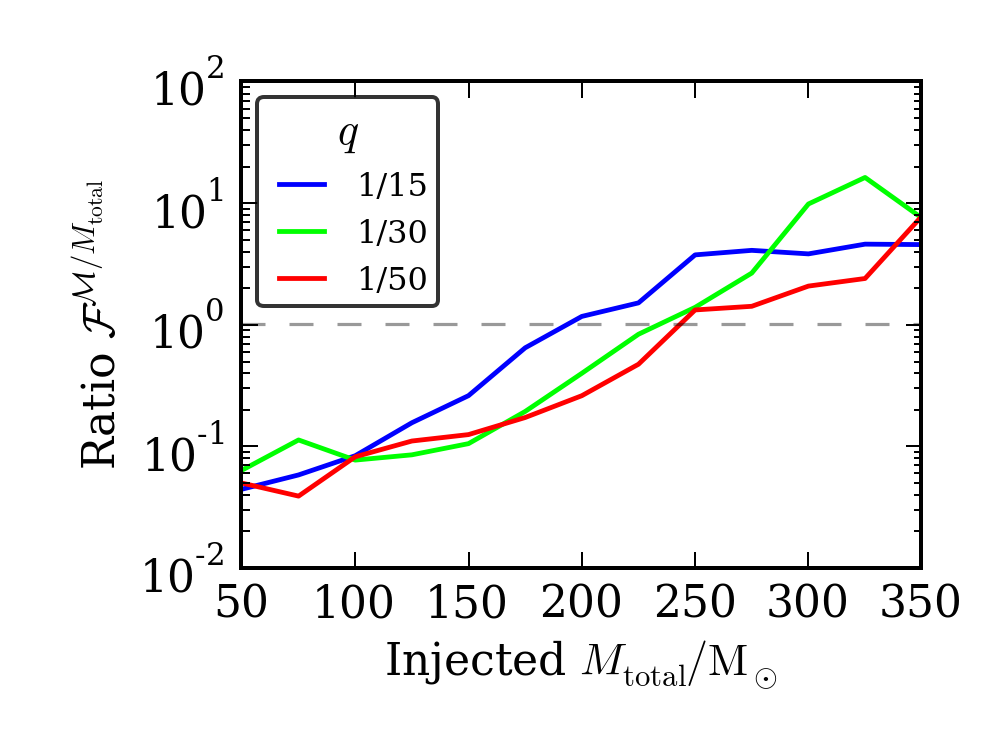} 
\caption{The ratio of fraction $90\%$ credible intervals (rescaled by injected values) for $\mathcal{M}$ and $\Mtot$, $\mathcal{F}^{\mathcal{M}/\Mtot} = (\mathrm{CI}_{0.9}^\mathcal{M}/\mathcal{M})/(\mathrm{CI}_{0.9}^{\Mtot}/\Mtot)$. The chirp mass is the better measured mass parameter for $\mathcal{F}^{\mathcal{M}/\Mtot} < 1$ and the total mass is better measured for $\mathcal{F}^{\mathcal{M}/\Mtot} > 1$; for $\Mtot \gtrsim 200~\Msun$, $\Mtot$ is the better measured mass parameter.}
\label{fig:McMtot_frac} 
\end{figure}

For low-mass systems, where the recovered SNR is dominated by the inspiral part of the coalescence \citep{Aasi:2013jjl}, the best constrained parameter is the chirp mass $\Mc=m_1^{3/5} m_2^{3/5} \Mtot^{-1/5}$. The uncertainty on the chirp-mass measurement is shown in \autoref{fig:Mc_bounds}. For greater $\Mtot$ the SNR becomes increasingly dominated by the merger and ringdown; the properties of the ringdown depend only on $\Mtot$ and the spin of the final BH $a\sub{f}$ \citep[cf.][]{Graff:2015bba,Veitch:2015ela}. High-mass systems, $\Mtot \gtrsim 200~\Msun$, are therefore better constrained in terms of their $\Mtot$ (\autoref{fig:Mtot_bounds}) than their $\Mc$, as shown in \autoref{fig:McMtot_frac}.
As discussed in \autoref{sec:flow}, since the measurement accuracy of $\Mc$ scales inversely with the number of in-band cycles of the inspiral, the specific mass of the $\Mc$--$\Mtot$ transition depends on the lower limit of the detector's sensitive frequency band $f\sub{low}$, either set explicitly as part of the analysis or implicitly by either a high noise floor or uncertain low-frequency calibration

\begin{figure*}
	\includegraphics[width=0.49\textwidth, keepaspectratio]{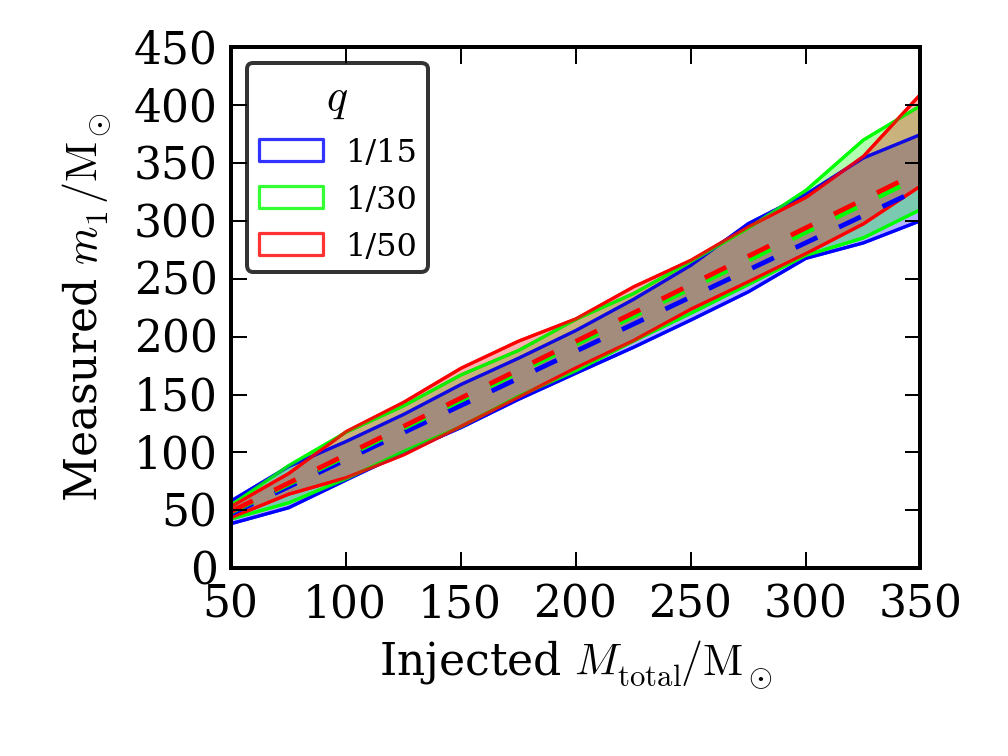}
	\includegraphics[width=0.49\textwidth,keepaspectratio]{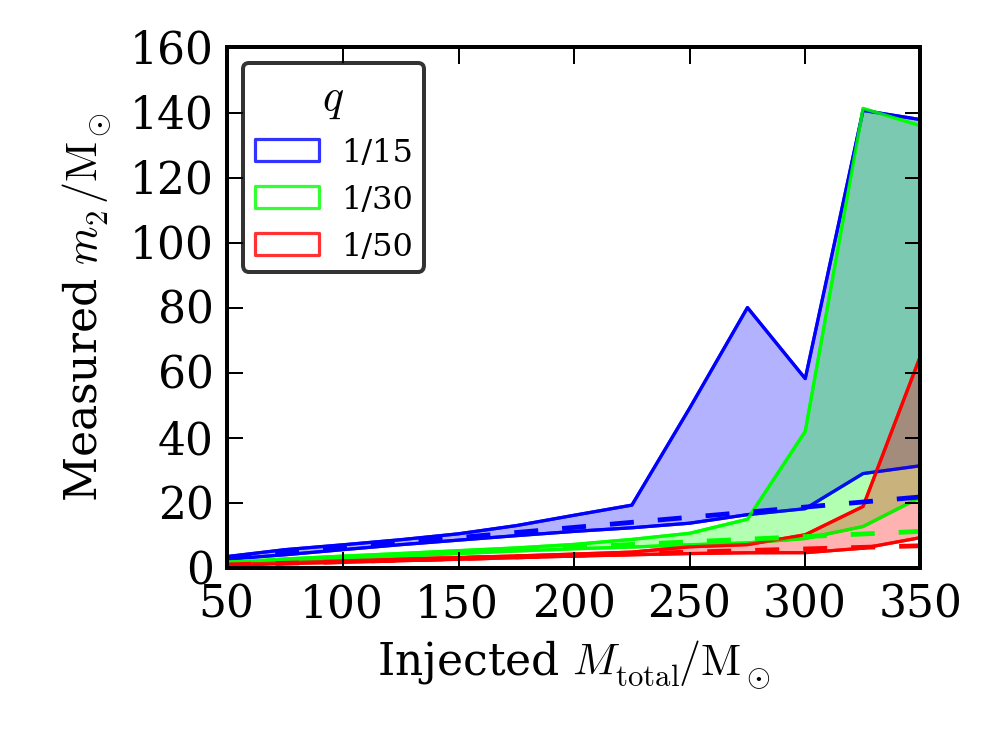}
	\caption{The 90\% credible intervals for the larger and smaller companion masses, $m_1$ (left) and $m_2$ (right), respectively. The mass of the secondary $m_2$ is poorly measured and biased upward (toward stronger signals which can be observed at greater distances) when $\Mtot$ is so large that little of the inspiral falls into the sensitive frequency band.}
	\label{fig:components}
\end{figure*}

The mass measurements can alternatively be represented by the $90\%$ credible intervals for the companion masses. \autoref{fig:components} shows that the larger mass $m_1$ is well constrained due to its near equivalence to $\Mtot$ for these systems. At low $\Mtot$, the mass ratio also provides tight constraints on $m_2$ compared to more equal mass systems \citep{Veitch:2015ela}. The strong dependence of the number of waveform cycles (in the detector band) upon the mass ratio means that even a small shift away from the true $q$ value causes a large dephasing between the signal and template waveforms (assuming that $\Mtot$ or $\Mc$ is well constrained), and therefore a rapid decrease in the measured likelihood. At high $\Mtot$ the detectors are only sensitive to the ringdown of the coalescence, where the mass-ratio dependence is only measured weakly through the final BH spin $a\sub{f}$.

\begin{figure}
\includegraphics{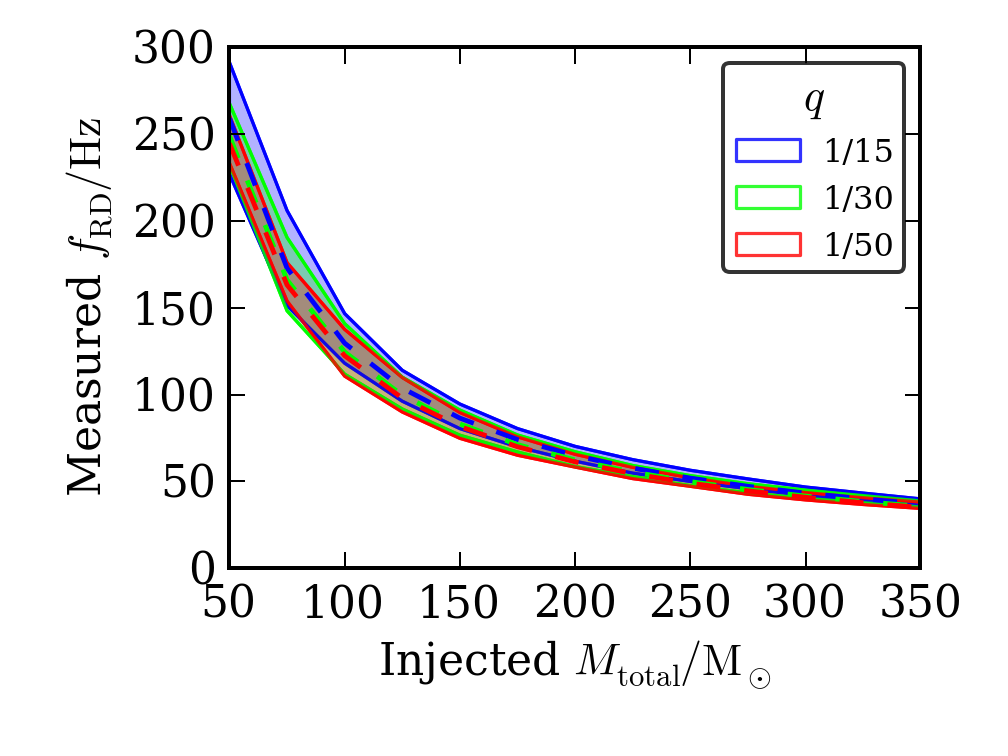}
\caption{The $90\%$ credible interval for $f\sub{RD}$. The true values are shown as dashed lines for each $q$. For high $\Mtot$ systems, whose in-band signal is dominated by the ringdown, $f\sub{RD}$ is better constrained than $\Mtot$.}
\label{fig:fRD_bounds} 
\end{figure}

The remnant spin together with the final mass $M\sub{f}$ determines the frequency of the quasi-normal modes (QNMs) of the ringdown of the merged BH. Following an approximate model for estimating $M\sub{f}$ and $a\sub{f}$ \citep[Equations 14 and 16]{Healy:2014yta} it is then possible to convert the posterior samples in the space of companion masses and spins into the frequency $f\sub{RD}$ of the $0$th overtone of the dominant $(l,m) = (2,2)$ QNM as \citep[Table \RNum{8}]{Berti:2005ys},
\begin{equation}
f\sub{RD} = \frac{c^3}{2\pi G M\sub{f}} {\left[0.5251 - 1.1568 (1 - a\sub{f})^{0.1292}\right]} \, .
\label{eq:fRD}
\end{equation}
\autoref{fig:fRD_bounds} compares the inferred $f\sub{RD}$ to the value of the ringdown frequency of the injected waveform. It is the most accurately measured parameter for high $\Mtot$ systems (cf.\ \citealt{Aasi:2014bqj} for IMBHB systems), while $\Mtot$ (cf.\ \autoref{fig:Mtot_bounds}) suffers from a partial degeneracy with spin.

\begin{figure}
\includegraphics{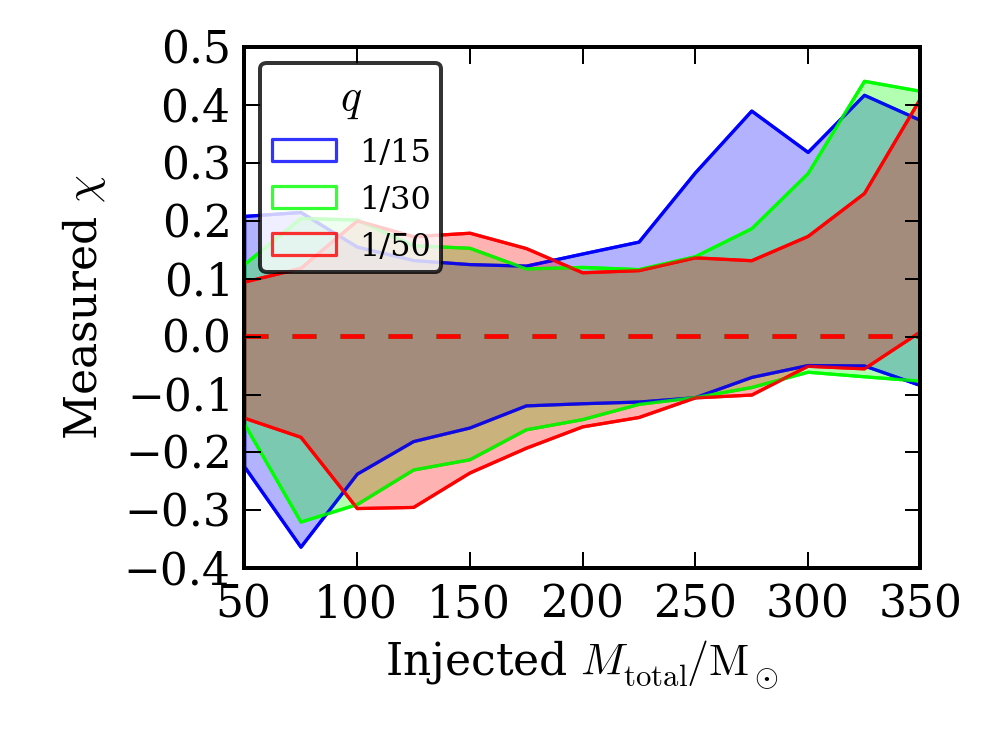} 
\caption{The $90\%$ credible interval for the effective dimensionless spin $\chi$, with all injections at $\chi=0$. Estimates are constrained to $\sim 1/5$ of the prior range for all $\Mtot$ and $q$. }
\label{fig:chi_bounds} 
\end{figure}

For these non-spinning injections, \autoref{fig:chi_bounds} shows the recovery of the combined effective spin $\chi \equiv (m_1 a_1 + m_2 a_2) / \Mtot$ \citep{Santamaria:2010yb}; $\chi$ encompasses both the relatively well measured spin of the higher mass companion and the unconstrained spin of the lower mass companion. The effective spin can be constrained to $\sim 1/5$ of the prior range, always being consistent with $\chi=0$. The trend towards larger positive $\chi$ for high $\Mtot$ is a consequence of the degeneracy between $\chi$ and $\Mtot$. There is a preference for systems at larger luminosity distances (as a consequence of our assumption of sources being uniformly distributed in volume), which makes signals quieter, but this can be compensated by an overestimation of $\Mtot$. This, combined with $q$ tending towards equal mass for those systems, forces $\chi$ to higher positive values in order to correct for the in-band length of the observed signal, the end of the inspiral, and the well measured ringdown frequency (cf.\ \autoref{eq:fRD}).

\subsection{Effects of cosmology on inferring the presence of an IMBH} 
\label{sec:Cosmo}

GW observations of IMRACs or IMBHBs may provide the first conclusive evidence for the existence of IMBHs. In order to infer the presence of an IMBH in an IMRAC, we need to be able to claim that $m_1$ is greater than a fiducial threshold $M_\mathrm{IMBH}$ at a desired confidence level. Here, we follow \citet{Veitch:2015ela} and adopt a threshold mass $M_\mathrm{IMBH} \geq 100~\Msun$.\footnote{\citet{Veitch:2015ela} found that $m_1 \gtrsim 130~\Msun$ was required to infer the presence of an IMBH in an IMBHB at $95\%$ confidence.} 

To infer the presence of an IMBH we must constrain the \emph{physical} mass of the source. GWs are redshifted due to the expansion of the universe. This corresponds to a redshifting of the companion masses as $m_{1,2} = m_{1,2}^\mathrm{source} (1 + z)$ for a signal at redshift $z$; thus far in the paper we have considered the redshifted masses as measured in the detector rest frame. Advanced ground-based detectors can observe IMRACs at maximum redshifts of $z \sim 0.2$--$1$, depending on the mass ratio \citep[e.g.,][]{Belczynski:2014iua}. A signal detected with $m_1 = 100~\Msun$ and redshift $z = 0.2$ would correspond to a physical system with $m_{1}^\mathrm{source} = 100/(1+0.2) \approx 83~\Msun$, which would not be an IMBH by our definition. It is thus necessary to fold in redshift information in order to produce robust IMRAC mass measurements.

For each of our systems, we obtain a posterior on the luminosity distance $D\sub{L}$. The luminosity distance is related to the redshift by \citep[section 15.8]{Hobson:2006}
\begin{equation}
\label{eq:luminositydistance}
D\sub{L}(z) = \frac{c (1+z)}{H_0} \int_{0}^{z} \frac{\dd z'}{\zeta(z')} ,
\end{equation}
where, if we assume zero curvature and neglect radiation energy density,
\begin{equation}
\zeta(z) = \sqrt{(1 + z)^{3} \Omega\sub{M} + \Omega_{\Lambda}} .
\end{equation}
We invert Equation~\ref{eq:luminositydistance} numerically to find $z(D\sub{L})$, adopting standard cosmological parameters: $\Omega\sub{M} = 0.3$, $\Omega_\Lambda = 0.7$ and $H_0 = 70.0~\mathrm{km\,{s}^{-1}\,{Mpc}^{-1}}$. We then calculate the primary mass in the source frame as
\begin{equation}
m_{1}^\mathrm{source} = \frac{m_{1}}{1 + z} .
\end{equation}

\begin{figure}
\includegraphics[width=0.5\textwidth]{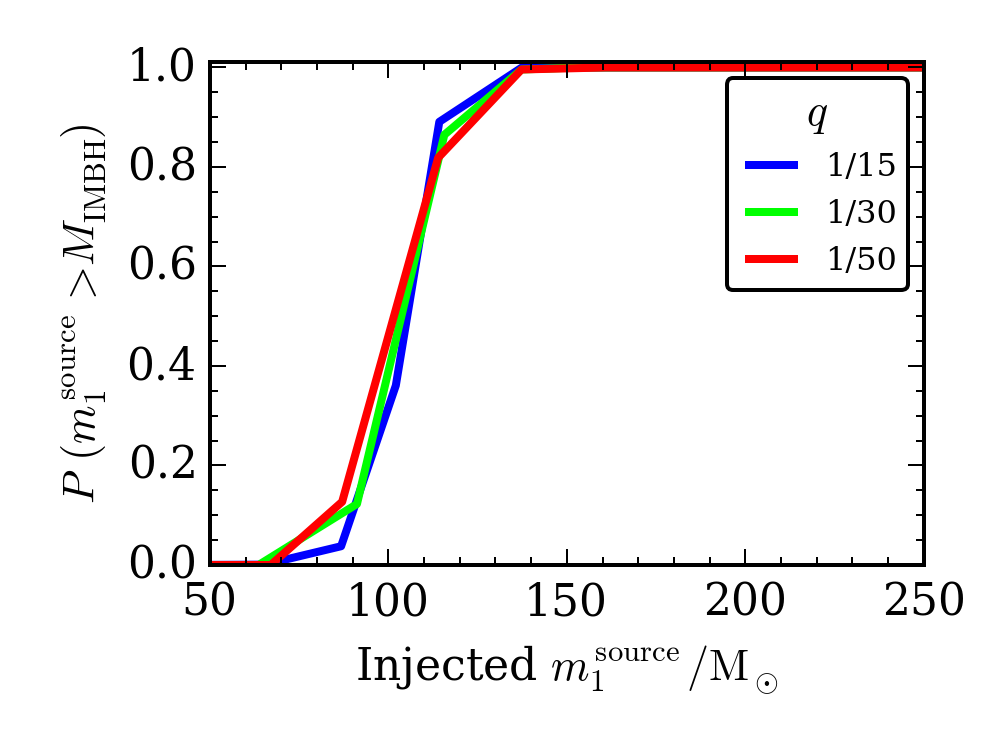} 
\caption{Fraction of the posterior distribution for $m_1\super{source} > M\sub{IMBH} \equiv 100\ M_\odot$ as a function of the injected primary source mass. The three curves correspond to the three different mass ratios considered. We find that we can infer the presence of an IMBH at 95\% confidence when the system has $m_1\super{source} \gtrsim 130~\Msun$.}
\label{fig:prob_m1_over_100} 
\end{figure}

In Figure~\ref{fig:prob_m1_over_100} we show the fraction of the posterior distribution for $m_{1}^\mathrm{source} > M_\mathrm{IMBH}$ as a function of the injected primary source mass. We find that we can infer the presence of an IMBH with mass above $100~\Msun$ at $95\%$ confidence when the system has $m_1\super{source} \gtrsim 130~\Msun$, matching \citet{Veitch:2015ela}. Additionally we can infer the presence of an IMBH at $\sim100\%$ confidence for systems with $m_{1}^\mathrm{source} \gtrsim 150~\Msun$.

\section{Discussion}
\label{sec:discussion}

Having completed our PE study, validating our ability to measure the mass and spin parameters of IMRAC systems, we now focus on the sensitivity and robustness of our results to a selection of assumptions adopted in our analysis: the low-frequency sensitivity of the detectors (\autoref{sec:flow}), the SNR of the signal (\autoref{sec:SNR}) and the accuracy of the waveform model (\autoref{sec:systematics}).

\subsection{Impact of low-frequency sensitivity}
\label{sec:flow}

As a consequence of the typical high total masses of IMRACs, the low-frequency sensitivity of the detectors is expected to be crucial for PE. IMRAC parameters are most precisely measured when they are determined by an inspiral with many in-band cycles, ending at the innermost stable circular orbit at a frequency $f\sub{ISCO}$. The transition in measurement accuracy seen in \autoref{fig:McMtot_frac} will therefore be shifted to lower masses for decreased low-frequency sensitivity, caused by either a high noise floor or uncertain low-frequency calibration.

\begin{figure}
\includegraphics[width=0.49\textwidth, keepaspectratio]{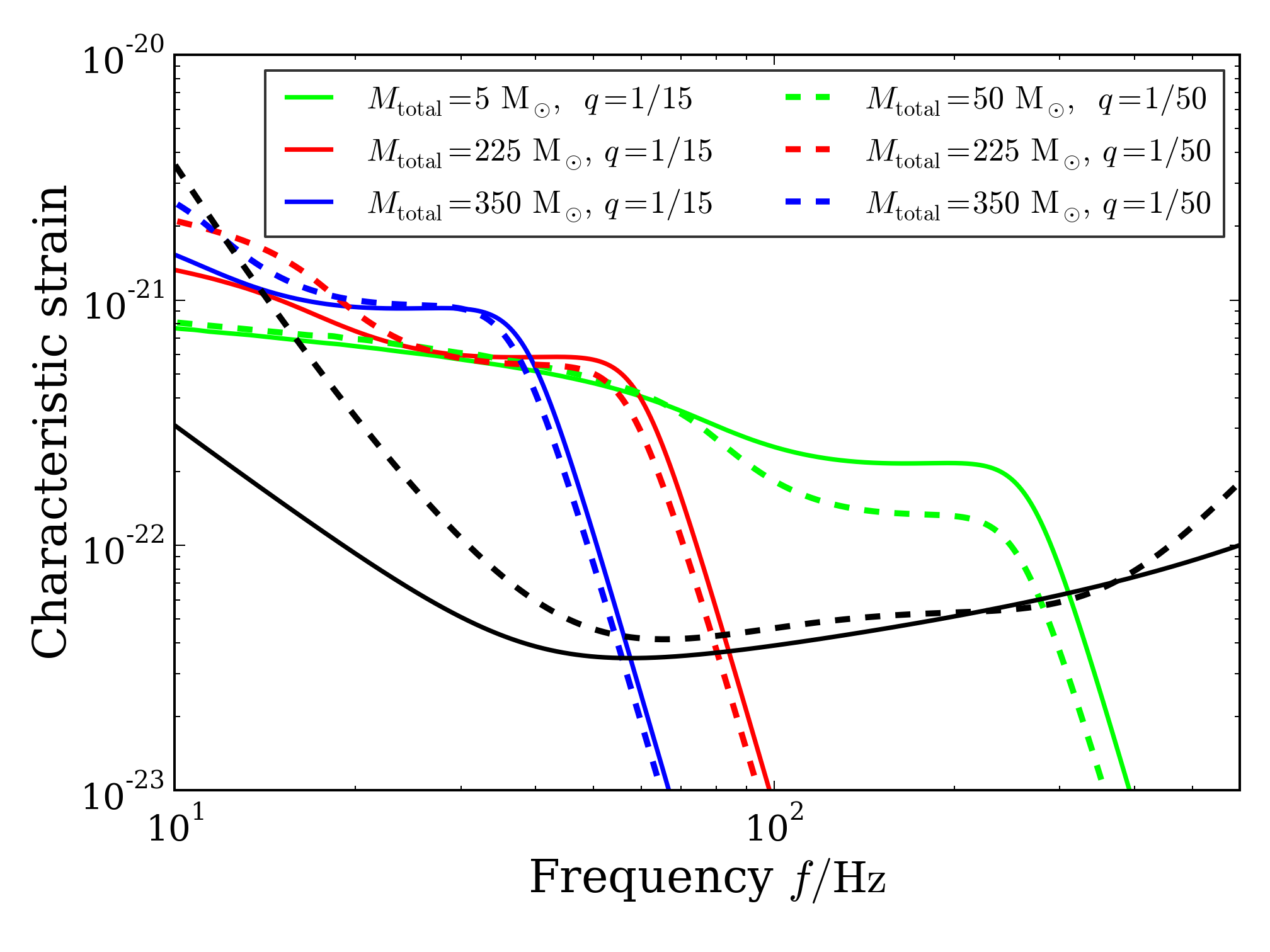} 
\caption{Characteristic strain $h_\mathrm{c} \equiv 2f \lvert \tilde h(f) \rvert$ \citep{Moore:2014lga} of \mbox{SEOBNRv2\_ROM\_DoubleSpin} for a range of the injected waveforms used in \autoref{sec:key_results}, all at $\rho=15$. In black, the dimensionless detector noise amplitude $h_n$ \citep{Moore:2014lga} of the aLIGO (solid) and AdV (dashed) design noise spectra.} 
\label{fig:wf_and_ASD} 
\end{figure}

\begin{figure}
\includegraphics[width=0.49\textwidth, keepaspectratio]{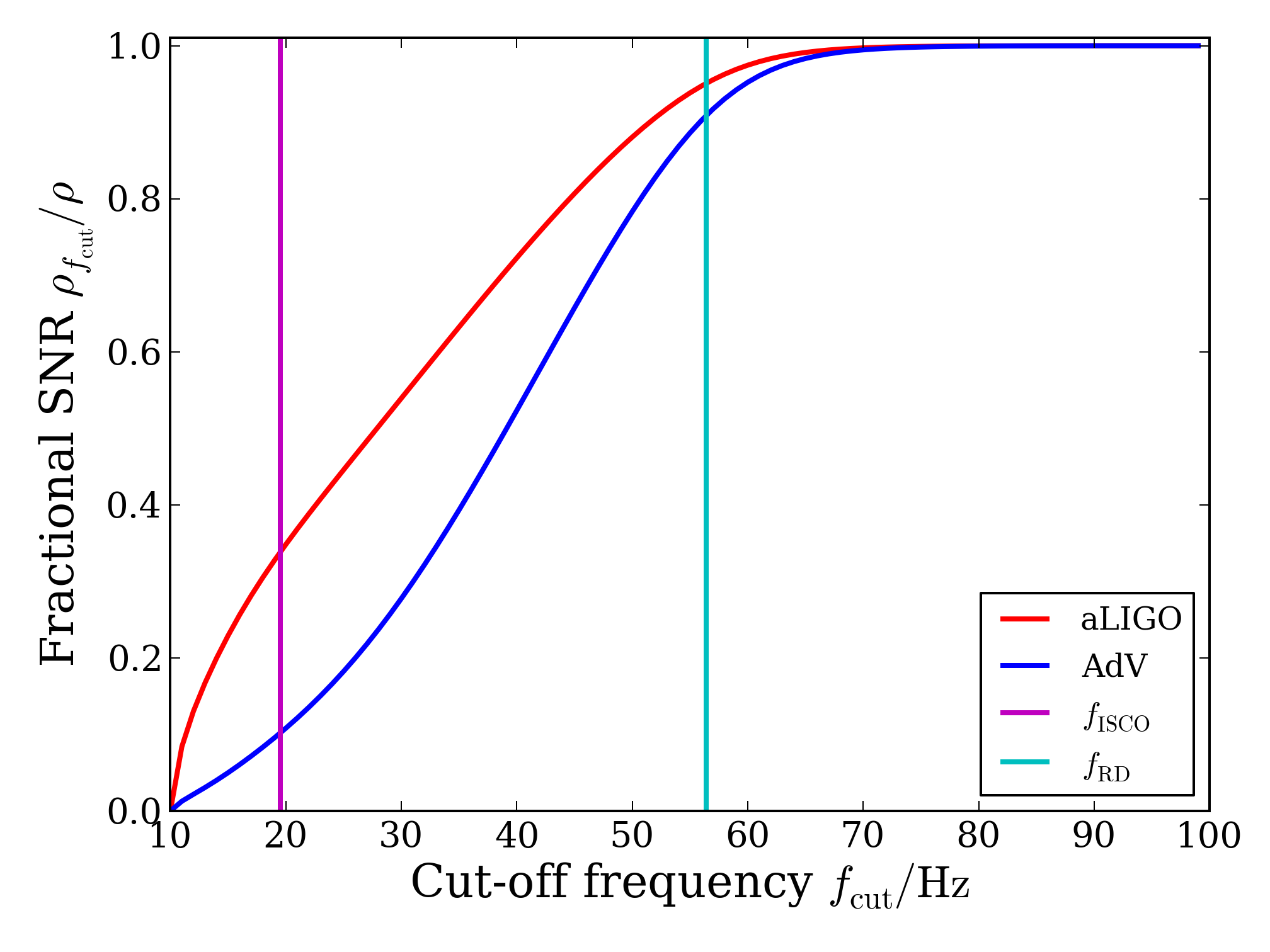} 
\caption{The SNR accumulated between $10\mathrm{Hz}$ and $f\sub{cut}$ $\rho_{f_\mathrm{cut}}$ as a fraction of the total SNR $\rho$, for a system with $\Mtot=225~\Msun$ and $q=1/15$ injected at $\rho=15$. This illustrates the relative low-frequency sensitivity between aLIGO and AdV used in this analysis, cf.\ \autoref{fig:wf_and_ASD}.}
\label{fig:flow_fraction} 
\end{figure}

\autoref{fig:wf_and_ASD} shows a selection of frequency-domain IMRAC waveforms and the detector noise curves, represented as characteristic strains and noise amplitudes respectively \citep{Moore:2014lga}. The randomly chosen sky locations and orientations of our injections mean that for some mock events, the majority of the network SNR is contributed by AdV, with its relatively poorer low-frequency sensitivity, illustrated in \autoref{fig:flow_fraction}. An example of this effect is seen in the $\Mtot=275~\Msun,\ q=1/15$ event clearly visible in \autoref{fig:Mc_bounds}. 

Low-frequency sensitivity is particularly critical for measuring the mass ratio, as the ringdown can only provide information on the total remnant mass and spin. For example, for a $\Mtot=225~\Msun,\ q=1/15$ system which sits at the transition of inspiral detectability with the aLIGO noise spectrum with sensitivity starting at $f\sub{low} = 10~\mathrm{Hz}$, the $90\%$ credible interval is $\mathrm{CI}_{0.9}^q \lesssim 0.05$. However, if sensitivity below $20~\mathrm{Hz}$ is lost, $\mathrm{CI}_{0.9}^q$ spans half of the allowed range from $0$ to $1$, although more than $90\%$ of the SNR is still available for detection (see \autoref{fig:flow_fraction}). If $f\sub{low}$ increases to $30~\mathrm{Hz}$ or above, $q$ becomes essentially unconstrained.

\subsection{Uncertainty versus signal-to-noise ratio}
\label{sec:SNR}

\begin{figure}
\includegraphics[width=0.49\textwidth, keepaspectratio]{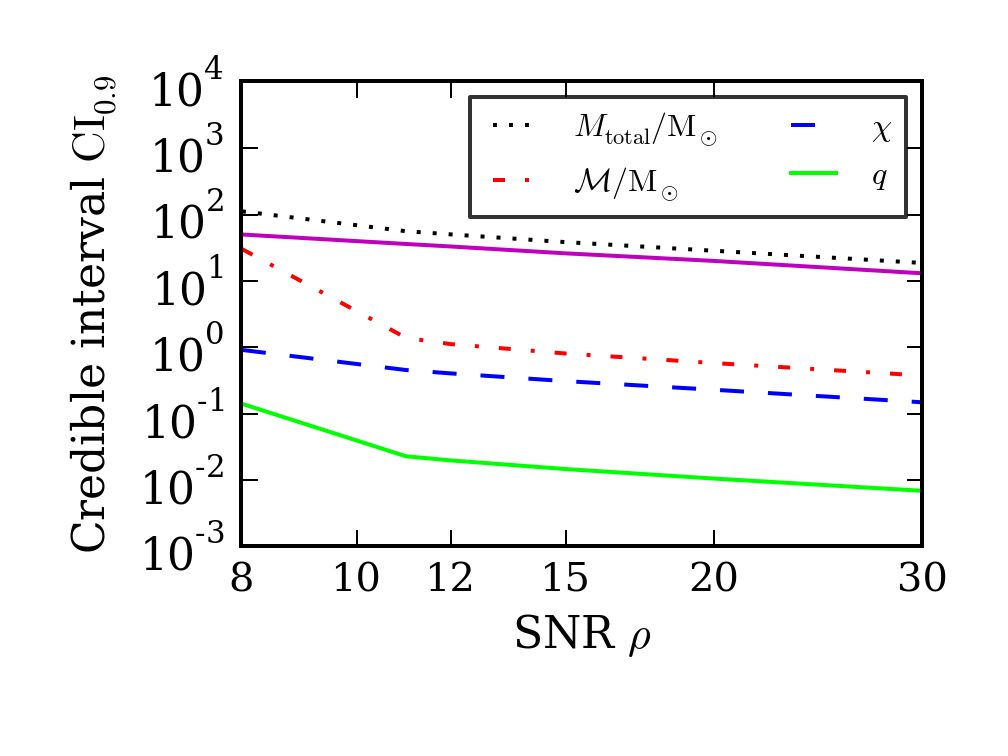} 
\caption{The width of the $90\%$ credible intervals as a function of injected network SNR $\rho$ for an exaple system ($\Mtot=155~\Msun$, $q=1/30$), sampled at the indicated $\rho$. At high $\rho$ this follows a $1/\rho$ trend, the slope of which can be gauged by comparison with the magenta line.}
\label{fig:SNR} 
\end{figure}

To investigate the effect of the loudness of the detected signal, a series of simulations at a range of SNRs were performed. For high $\rho$ the shape of the posterior distribution approaches a multivariate Gaussian, and thus the uncertainties on individual parameters scale as $\rho^{-1}$ \citep[cf.][]{Vallisneri:2007ev}. As shown in \autoref{fig:SNR}, this behaviour can be observed starting at $\rho \sim 11$. Hence, it should be possible to scale our results to estimate PE ability for other detectable signals.

\subsection{Systematics}
\label{sec:systematics}

At the high mass ratios of IMRACs, the post-Newtonian expansion breaks down, an extreme-mass-ratio expansion in the mass ratio \citep[the self-force problem;][]{Poisson:2011nh} is not yet sufficient \citep[e.g.,][]{Mandel:2008bc}, and numerical-relativity solutions are extremely computationally expensive \citep[e.g.,][]{Lousto:2010tb,Husa:2015iqa}. 
Therefore, possible model errors and the ensuing systematic bias in parameter recovery are a significant concern \citep{Smith:2013mfa}. To validate our choice of SEOBNRv2 for this study, a subset of the events shown in \autoref{sec:key_results} were repeated as injections with a different waveform family, but still recovered with \mbox{SEOBNRv2\_ROM\_DoubleSpin}. The injections were performed with IMRPhenomD \citep{Husa:2015iqa, Khan:2015jqa}, a phenomenological waveform model constructed in the frequency domain and calibrated against numerical relativity up to mass ratios $q \geq 1/18$.\footnote{In IMRPhenomD, IMR refers to inspiral--merger--ringdown, not intermediate mass ratio.}

\begin{figure*}
	\includegraphics[width=0.49\textwidth, keepaspectratio]{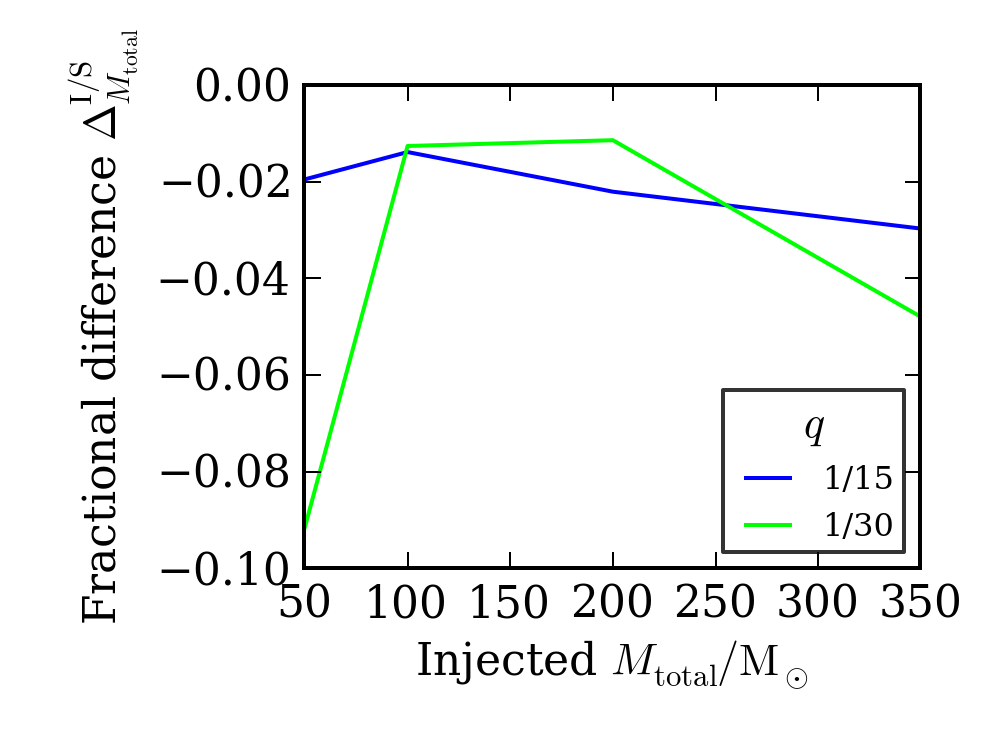}
	\includegraphics[width=0.49\textwidth,keepaspectratio]{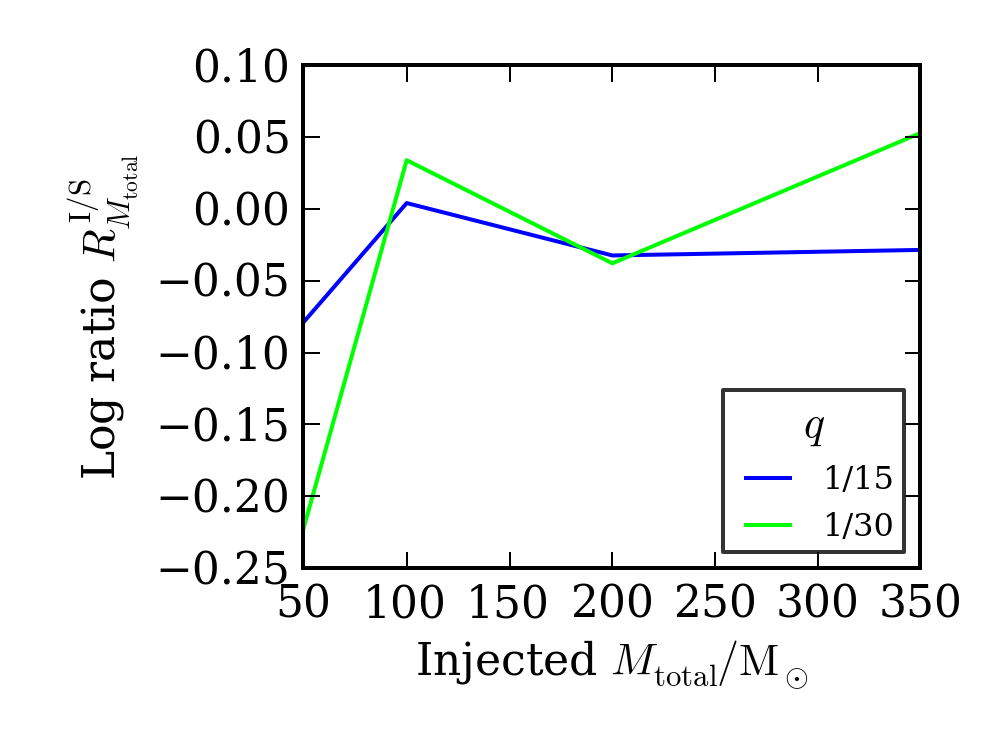}
	\caption{Comparison of results recovered for IMRPhenomD (I) and SEOBNRv2 (S) injections. \mbox{SEOBNRv2\_ROM\_DoubleSpin} templates are used for PE in both cases. The left panel shows the fractional difference in the recovered means of the $\Mtot$ posterior distributions $\Delta_{M_\mathrm{total}}^\mathrm{I/S} = \left(\langle{M_\mathrm{total}}\rangle^\mathrm{I} - \langle{M_\mathrm{total}}\rangle^\mathrm{S}\right)/ \langle{M_\mathrm{total}}\rangle^\mathrm{S}$ as a function of $\Mtot$ and $q$. If the two injections were recovered with identical posterior means $\Delta_{M_\mathrm{total}}^\mathrm{I/S} =0$. The right panel shows the natural logarithm of the ratio of $\Mtot$ $90\%$ credible intervals $R_{M_\mathrm{total}}^\mathrm{I/S} = \ln\left({\mathrm{CI}_{0.9,\,\mathrm{I}}^{M_\mathrm{total}}}/{\mathrm{CI}_{0.9,\,\mathrm{S}}^{M_\mathrm{total}}}\right)$. If the posteriors have the same width, then $R_{M_\mathrm{total}}^\mathrm{I/S} = 0$; $R_{M_\mathrm{total}}^\mathrm{I/S} < 0$ indicates that the posterior distribution is narrow for the IMRPhenomD injection than for the SEOBNRv2 injection.} 
	\label{fig:Mtot_systematics}
\end{figure*}

The systematic bias in the recovered $\Mtot$ introduced by the difference between IMRPhenomD injections and \mbox{SEOBNRv2\_ROM\_DoubleSpin} templates is small for all investigated systems, as shown in \autoref{fig:Mtot_systematics}, confirming the results shown in \citet{Khan:2015jqa}. Additionally, $\mathrm{CI}_{0.9}^{\Mtot}$ remains largely unaffected, even outside the region where IMRPhenomD has been calibrated to numerical-relativity waveforms. 

Using the same condition as in \autoref{sec:Cosmo} for determining the presence of an IMBH, we find that the threshold of $m_1\super{source} \gtrsim 130~\Msun$ is robust under this systematic bias. Therefore, we expect that, if the difference between IMRPhenomD and SEOBNRv2 is typical of the waveform model uncertainty in the IMRAC space, then the systematic error introduced from waveform uncertainty should not hinder the identification of IMBHs from IMRAC observations. We evaluated the impact of systematics by comparing posterior probability distributions computed with different models of spin-aligned, circular templates without higher-order modes; the impact of systematics will need to be re-evaluated once waveforms incorporating all of these effects are available.

\section{Summary}
\label{sec:summary}

IMBHs may play an important role in the formation of supermassive BHs, and the dynamics of dense stellar environments like globular clusters \citep[e.g.,][]{Trenti:2006cp,Gill:2008pt}, yet conclusive evidence for their existence remains elusive. 
A network of advanced GW detectors can observe an IMBH as part of an IMRAC at cosmological distances, out to redshift $z\gtrsim 0.5$. Recent progress in the development of waveforms suitable for IMRAC systems has enabled the first systematic study of the measurement of the masses and spins of IMRACs. Despite the short in-band signal, we find that inference on the emitted GWs can provide interesting measurements of IMRAC systems. 

For low mass IMRAC systems, $\Mc$ is the best constrained parameter. As total mass increases, the inspiral moves out of the sensitive frequency band of the detectors, after which most of the information come from the ringdown of the merger remnant, so that the ringdown frequency is best constrained. For high-mass systems, $\Mtot$ is measured more precisely than $\Mc$, but is still partly degenerate with the (poorly constrained) spin.

With a low and stable noise floor at low frequencies, it will be possible to infer the presence of an IMBH with mass $\geq 100~\Msun$ at $95\%$ confidence for systems with $m_1\super{source} \gtrsim 130~\Msun$. This relies on the assumption of standard cosmology to infer the source mass from the measured redshifted mass and luminosity distance.

By using different waveform approximants for signal injection and PE, we confirm that our results, including the detectability of an IMBH, are robust against potential waveform errors so long as they fall in the range bracketed by these approximants.

Future investigations of IMRACs will benefit from ongoing waveform development to include spinning and precessing signals, possibly eccentric binaries, and higher-order modes. Building upon improved confidence in parameter estimation with IMRAC waveforms, future studies could focus on using IMRAC observations to enhance our understanding of globular-cluster dynamics, and the suitability of IMRACs for high-precision tests of general relativity in the strong-field regime \citep[e.g.,][]{Brown:2006pj,Gair:2007kr,Rodriguez:2011aa}.

\section*{Acknowledgements}

We are especially grateful to Michael P\"urrer for assistance in using his ROM waveforms and comments on the manuscript. We thank Tom Callister and Tom Dent for suggestions and discussion.
This work was supported in part by the Science and Technology Facilities Council. JV was supported by STFC grant ST/K005014/1. IM acknowledges support from the Leverhulme Trust. 
LIGO was constructed by the California Institute of Technology and Massachusetts Institute of Technology with funding from the National Science Foundation and operates under cooperative agreement PHY-0757058. 
We are grateful for computational resources provided by the Leonard E.\ Parker Center for Gravitation, Cosmology and Astrophysics at University of Wisconsin--Milwaukee.
 This is LIGO document number P1500207.




\bibliographystyle{mnras}
\bibliography{imri_pe}






\bsp	
\label{lastpage}
\end{document}